# PCF Scheme for Periodic Data Transmission in Smart Metering Network with Cognitive Radio

Yue Yang, *Student Member IEEE*, Sumit Roy, *Fellow IEEE*

## Abstract

The next generation Advanced Metering Infrastructure (AMI), with the aid of two-way Smart Metering Network (SMN), is expected to support many advanced functions. In this work, we focus on the application of remote periodic energy consumption reporting, which is a fundamental and significant component of Demand Response and Load Management. In order to support this periodic application with satisfactory communication performance, a well-suited Media Access Control (MAC) protocol needs to be designed. Because the number of Smart Meters (communication nodes) involved in SMN are much larger than that in today's local area networks, the traditional taking-turns MAC protocol, such as Point Coordination Function (PCF) in WiFi is unlikely to perform well. In order to solve this problem, we propose a modified PCF scheme with the combination of Cognitive Radio technology, in which the Smart Meters may use the free channels (white space) to report energy consumption data to the Local Collector when the Primary Users are not occupying the channels. We also conduct comprehensive throughput analysis on the proposed scheme. The numerical results and simulation results through NS-3 show that the PCF scheme with Cognitive Radio significantly outperform the traditional one in a densely populated network like SMN.

## I. Introduction

Traditionally, the collection of power consumption data from the end-customer premises has been accomplished by using conventional meters. Even if such meters could be remotely read as in Automated Meter Reading (AMR) system, such capabilities were limited to infrequent one-way data upload. The next generation power metering system - Advanced Metering Infrastructure (AMI) - is based on two-way communication transceivers integrated into the Smart Meter. Therefore, besides the remote reading function, AMI may also support load management and demand response by sending the down-link control information from the utility.

Since AMI crucially depends on a two-way Smart Metering Network (SMN), SMN design considerations have begun to draw attentions. The pros/cons of several optional bi-directional communication technologies and sensing technologies appear in [1], [3], [4], [17], [18]; unlicensed wireless technologies, such as 802.11 Wireless LANs, appear to be preferred options, for their low cost and widespread prior adoption. Furthermore, in order to solve some potential challenges unique to SMN scenarios, a new IEEE Standardization Task Group (IEEE 802.11ah TG) [5] is currently engaged in creating a modified 802.11 based standard that operates at the frequencies below 1 GHz.

A critical choice for SMN is a suitable Multiple Access Control (MAC) protocol. The SMN communication may be classified according to their traffic types, such as periodic and event-driven; thus each type may require a different MAC protocol which is specially designed for it. For example, as an important part of the demand response and load management applications, every Smart Meter needs to report the energy consumptions and other related parameters to the utility once every interval, such as 15 mins [2]. Furthermore, the detailed data format of the periodic energy consumptions report is also defined in the Decade 3 of Standard ANSI C12.19 [8]. After that, according to [10], the utilities would exploit these periodic data to analyze the power consumption situation over the entire grid and conduct the load management when necessary. Since the energy consumption data reporting is periodic, it is a straightforward mechanism to use the centralized control and taking-turns MAC protocol to support the communication of such data, with the advantage of guaranteeing a certain rate for every Smart Meter [2]. For instance, the Point Coordination Function (PCF) defined in WiFi is such a suitable MAC protocol, in which the Access Point (AP) polls each station one by one and schedule their transmissions without any contention. Furthermore, the PCF based MAC protocol is also accepted by the IEEE 802.11ah TG when considering the collection of periodic data from Smart Meters at Local Collector. However, the traditional PCF protocol does not work well in SMN due to the scalability issue. As mentioned in [5], [9], both the coverage of network and the number of Smart Meters involved in the SMN significantly exceeds those in the indoor Wireless LAN (WiFi) networks for which they were originally designed. For example, IEEE 802.11ah TG requires one Local Collector to support a network with upto 6000 Smart Meters as benchmark, while the average number of communication nodes served by a single WiFi BSSID is of the order of 20-30. As the number of communication nodes is increased, the throughput will get saturated. As a result, the total time for the Local Collector to receive all the periodic data from the Smart Meters will be linearly increased. Therefore, in this work, we aim to design a PCF based MAC protocol for the periodic traffic in SMN, such as energy consumption report, and against the impact of scalability.

Fortunately, the concept of Cognitive Radio (CR) gives us an option to solve this problem. Currently, it has been more and more attractive that the CR is included in the Smart Grid Communication, especially in SMN. In brief, the communication node implemented with cognitive radio is able to scan and identify some available channels which are not occupied by the Primary Users (PUs). After that, they can switch their transceivers to the identified available channels and communicate with other nodes through these channels. Exploiting this special function of CR, we propose a modified PCF MAC protocol combined with CR, in which the Local Collector with multiple antennas is able to scan and identify all the available channels. After that, the collector will allocate all those identified channels to its associated Smart Meters and poll them by broadcasting one piece of control message. Then all the polled Smart Meters will reply the message and send their energy consumption data to the Local Collector simultaneously. Because all the channels are mutually independent and the collector is implemented with multiple antennas, the parallel transmissions of reporting data are able to be received by the collector at the same time and without any contention. In sum, we summarize four main benefits obtained from the combination of CR and PCF:

- The inclusion of cognitive radio helps the system to achieve spectrum efficiency and mitigate spectrum scarcity [11]. According to [5], the IEEE 802.11ah standard defines several schemes of channelization. With the aid of CR, we may employ the channelization with small bandwidth and thus support larger capacity without any sacrifice of performance.
- With the aid of CR, we may employ more channels and achieve parallel transmissions, which reduce the overall time to finish all the transmissions in the report period.
- The usage of CR can strengthen the robustness of the SMN. As mentioned in [12], [13], more available channels provide Smart Meters with backup channels to communicate with the Local Collector when the performance of dedicated channel is not good enough.
- According to [14], [15], the generated energy-related uplink data from Smart Meters will be up to tens of thousands of terabytes per year per utility. This poses a significant challenge for any existing communication network to collect, transmit, and store such large-scale data, especially for SMN with low data rate. The usage of CR potentially improves spectrum utilization and communication capacity to support large-scale data transmissions.

In the rest of the paper, we give an overview of the AMI Communication System Architectures and the traditional PCF scheme in Section II. After that, we give a detailed description of the modified PCF MAC protocol with CR (CR-PCF) in the Section III. The numerical analysis on the performance of traditional PCF and CR-PCF is presented in the Section IV. In Section V, we compare the performance between these two schemes through both numerical results and simulation results (NS-3). The entire paper is concluded in the Section VI.

## II. Network Topology and Overview of PCF

### A. AMI Communication System Architectures

We first give a high-level architecture of the AMI system based on the Smart Grid Architecture in [6], [7] shown in Fig.1. A short overview of its main components follows:

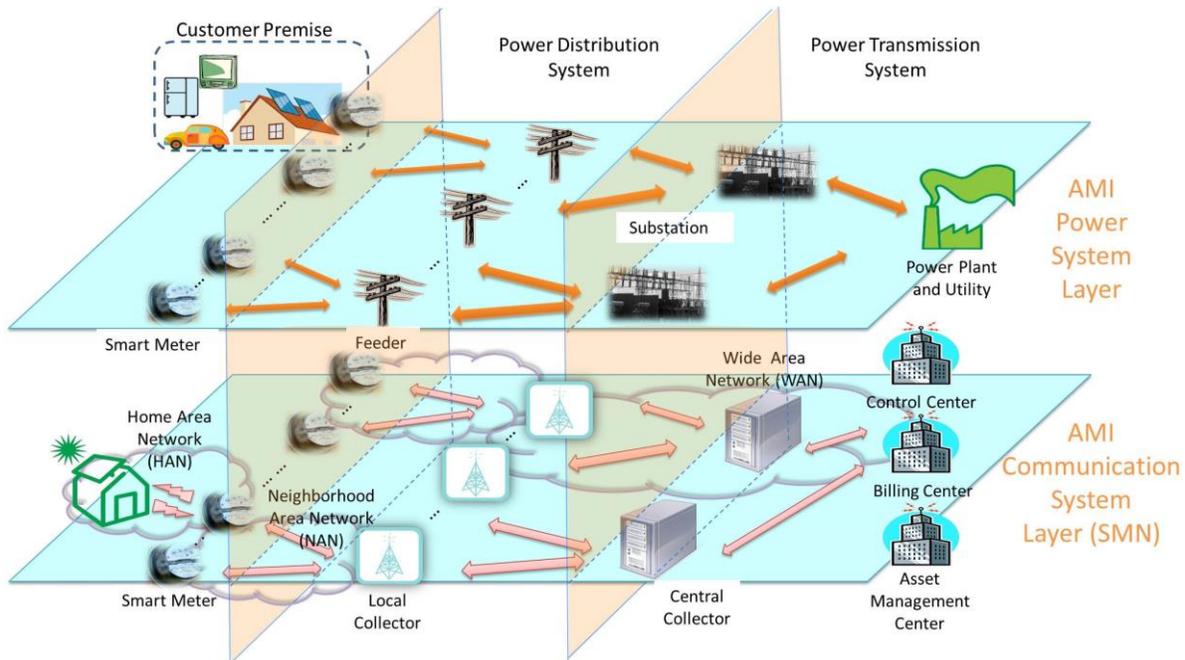

Fig. 1  AMI Power and Communication System Architectures

- Smart Meter (SM): This device integrated with two-way communication transceiver has three different roles. First, the SM is a multi-utility instrument measuring electric power consumption. It can thus act as an energy control center of the Home Area Network (HAN) that connects home appliances. Finally, the Smart Meter also serves as the gateway between HAN and external network; it reports on energy consumption, sends out urgent data, and receives remote commands from utility.
- Home Area Network (HAN): This is composed of multiple inter-connected electric appliances, such as air conditioner, plug-in hybrid electric vehicles (PHEV) etc. and the Smart Meter. All the components inside HAN share information or deliver control commands to each other.
- Local Collector (LC): Between the SM and Utility Center, there is a data collection node closer to customer premises - named as Local Collector - collects SM data and relays it to the Central Collector. Additional functions at the LC may include simple data processing and local decision making. The network segment between SM and LC is called Neighborhood Area Network (NAN), while that above LC belongs to Wide Area Network (WAN).

- Central Collector: A centralized data repository for the entire region operated by the utility - called Central Collector - acts as the interface with Control Center, Billing Center and Asset Management Center. Therefore, the three Centers may use SM data to conduct analysis and evaluate system status, and make decisions or deliver control commands to other components.

The HAN, consisting of electric appliances which are manufactured by different vendors, has a lot of flexibility in implementation. Furthermore, the design of WAN does not only depend on the communication requirements of SMN because it also includes some IEDs which serve the power systems other than AMI. Therefore, the utility operating SMN may leave the HAN and WAN open and put the focus of MAC protocols design on the segment of NAN, which is exactly the use case defined in the IEEE 802.11ah TG [16] and the network we analyze in this paper.

### B. Overview of Traditional PCF

Point Coordination Function (PCF) is defined in the legacy 802.11 Standards to support the time-bound and periodic communication services. For example, applying the PCF on the energy consumption reporting in NAN, the LC broadcasts a beacon at the beginning of the Contention Free Period (CFP). After that, the LC polls one SM and asks for the transmission of a packet. Upon receiving the poll message, the polled SM acknowledges it and sends back a pending packet. After LC receives this packet and checks its correctness, it acknowledges the successful data reception and piggybacks a piece of poll message to another SM. On the other hand, if the LC receives no response from the polled SM after waiting for a PCF Interframe Space (PIFS), it ignores the first SM and polls the next one. As shown in the Fig.2, the duration to finish one Poll-ACK message and payload transmission is defined as one Periodic Duration. This operation continues until all the energy consumption report packets are received at the LC side. PCF seems to work fine in this application when the number of SMs is limited. However, as shown in the Section V, the throughput of traditional PCF gets saturated as the number of the SMs increases. As a result, the total time to collect all the reporting packets at LC will be increased. In order to solve this problem, we combine the PCF scheme and cognitive radio technology to solve the scalability issue, which is presented in detail in the following section.

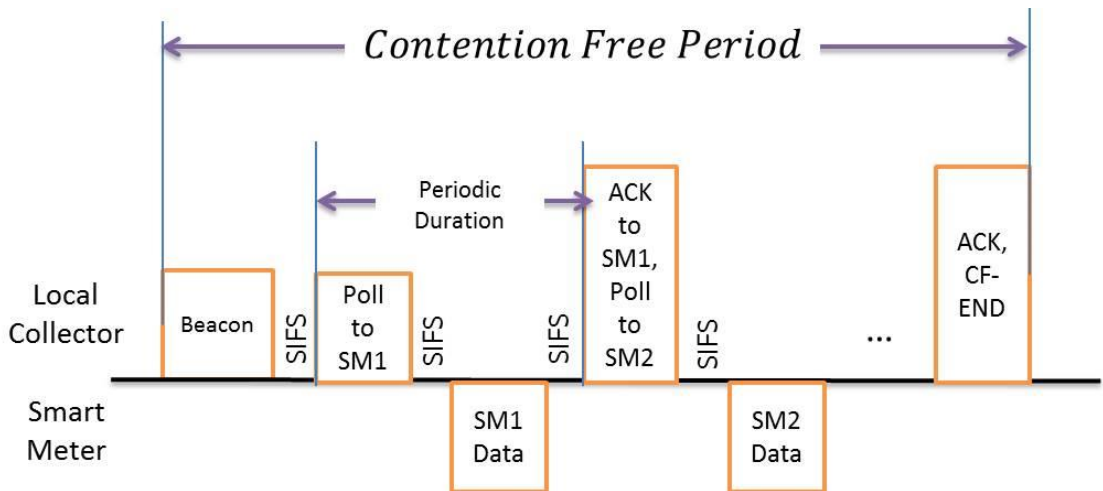

Fig.2 Traditional PCF Scheme Operation Example

# III. Modified PCF Scheme with Cognitive Radio

## A. System Model and Assumptions

Based on the topology of NAN, we make the following assumptions to formulate the system model.

- There are $M$ SMs, which are randomly distributed over the coverage of one LC.
- Since the uplink report data has a fixed format, as shown in ANSI C12.19 [8], the LC is able to know the number of packets to be transmitted from each SM, $L$, and the identical length of each packet $X$.
- In the NAN, there is one dedicated channel $H_0$, in which the LC and SM conduct reliable communications of payload packets or control message without the interference of PU.
- Besides the dedicated channel $H_0$, there are other $N$ potential available channels $\{H_i | i = 1, 2, \ldots, N\}$, in which the LC and SM work as the secondary users. It is noted that those $N$ channels are mutually independent and interference-free.
- The LC is implemented with $N + 1$ antennas, so that it is able to receive the packets from the $N + 1$ individual channels or sense these $N + 1$ channels simultaneously. However, the LC is not able to sense and communicate at the same channel at the same time.
- Each SM only needs to be implemented with one antenna, which can be switched among the $N + 1$ channels.
- We model the traffic behavior of PU on each of the $N$ supplementary channels as an ON-OFF state alternation, i.e. the PU communicates for a time $T_{ON}^1$, and then turns off and remains off for a time $T_{OFF}^1$. After that, it turns on for $T_{ON}^2$ and off for $T_{OFF}^2$ and so on. Furthermore, we assume that the $T_{ON}$ and $T_{OFF}$ follows the exponential distribution:

$$T_{ON} \sim exp(Z_{ON}) \text{ and } T_{OFF} \sim exp(Z_{OFF}).$$

Therefore, the long-run proportion of channel off is:

$$\gamma = \frac{Z_{OFF}}{Z_{OFF} + Z_{ON}}.$$

For simplicity, we consider the packet error only results from the collision between secondary user transmissions and primary user transmissions in this paper.

## B. Initialization

Based on the assumptions mentioned above, the LC, similar to the traditional PCF, broadcasts a beacon to initiate the CFP. After that, it initializes the first channel allocation and transmission according to the algorithm shown in the Fig.3.

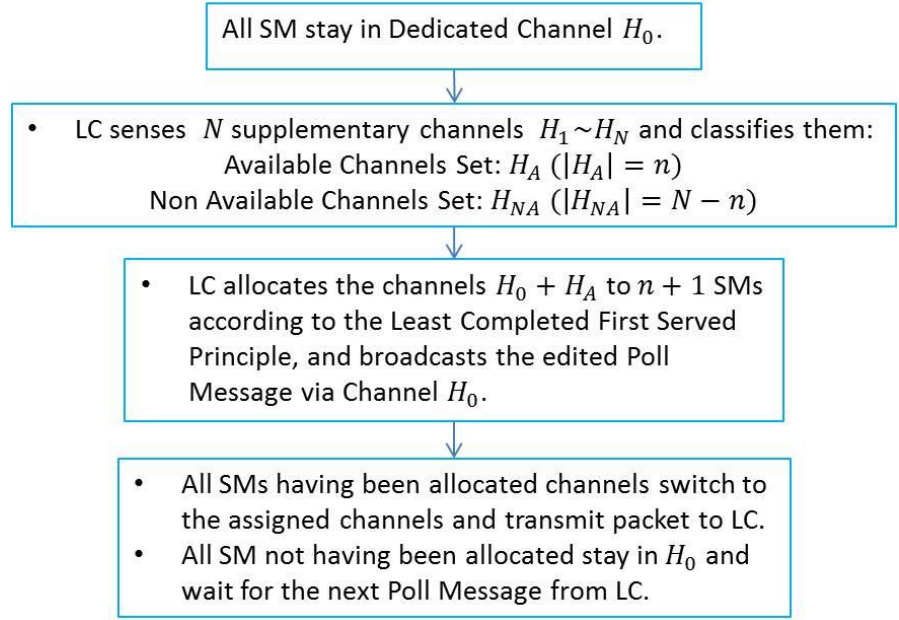

Fig.3　Initialization of CR-PCF Scheme

Compared to the Poll frame in the traditional PCF, the modification of that in our proposed protocol is quite limited. We only need to enclose the ACK and channel allocation information for all the SMs in the frame body. On the other hand, the fairness among all the SMs is also quite important in the protocol design. For example, the LC first serves the specific $n+1$ SMs and does not allocate channels to the rest of the SMs until the first $n+1$ SMs finish their transmission of all $L$ packets. After that, the LC focuses on serving another group of $n+1$ SMs until they finish their $L$-packet transmissions. As a consequence, there may exist a situation that the number of SMs which still need to be served is smaller than the number of available channels when approaching the end of CFP. It is obvious that this case is a waste of channel resource, which will also lead to increasing the duration to finish the entire reporting transmission. In order to avoid this situation, we propose a Least Completed First Served Principle as shown in the Fig.4. In brief, at each round of channel sensing and allocation, the LC prefers to allocate the channels to the SMs with the least completed transmissions so as to try to guarantee the fairness among all the SMs.

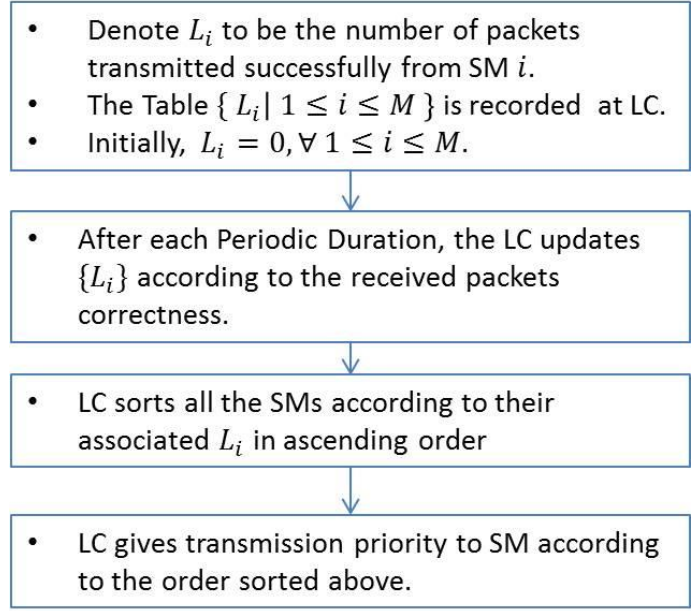

Fig.4  Least Completed First Served Principle

## C. Stepwise Algorithm

After the initial channel allocation and payload transmission, we design the stepwise algorithm according to which the LC and SM will work during each Periodic Duration. In this algorithm, only the LC is required to sense all the channels while the SMs only need to follow the scheduling and control message from LC. The details and operation example are shown in the Fig.5 and Fig.6 respectively. From the example, it is easy to find out that the gain of the throughput performance comes from the parallel utilization of available channels. However, the cost is that we have to extend the SIFS ahead of Poll-ACK Message to $\tau_{Sense}$ so as to guarantee the LC is able to sense all the available channels correctly enough. Furthermore, the SIFS right behind Poll-ACK Message is also required to be extended to $\tau_{Switch}$ so that the SMs having been allocated the available channels have enough time to switch to those channels. It is also noted that because $\tau_{Sense} > \tau_{Switch}$, all the SMs are able to switch back to $H_0$ as the LC is sensing the channels.

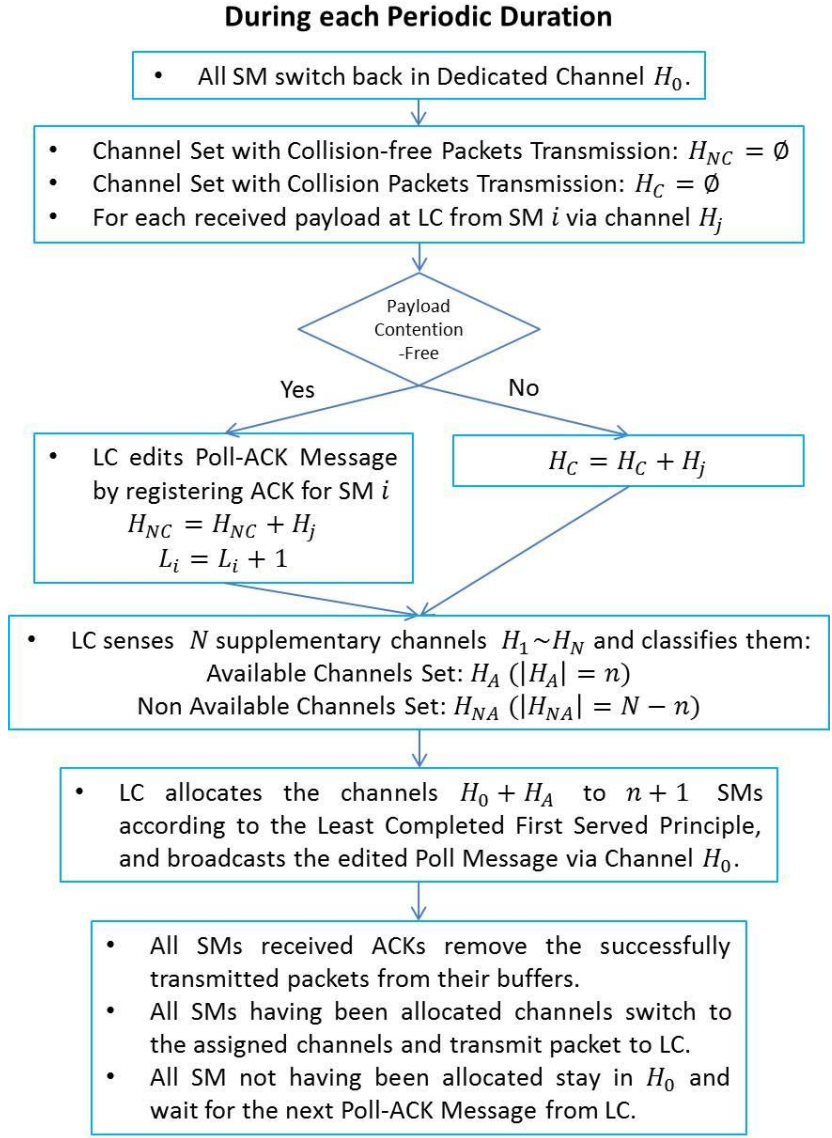

Fig.5　Stepwise Algorithm of CR-PCF Scheme

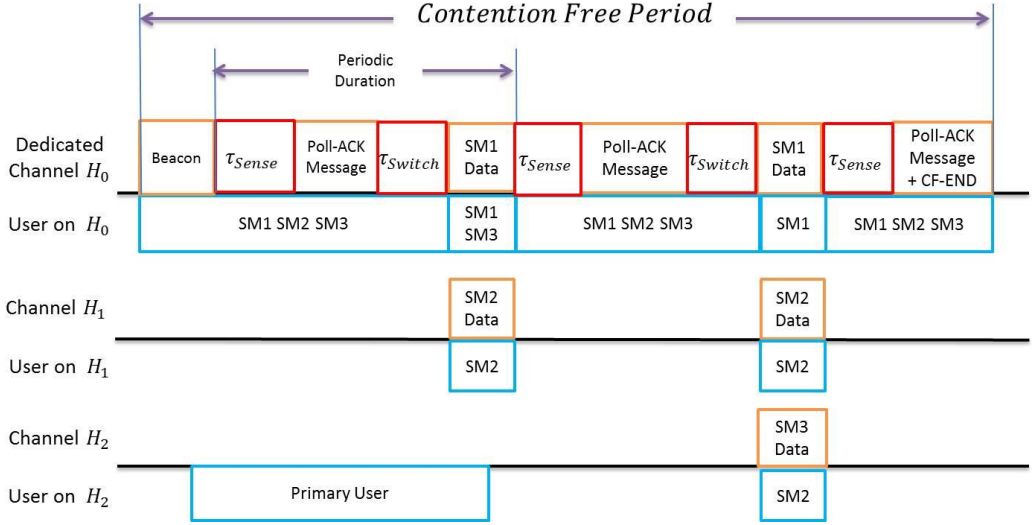

Fig.6   CR-PCF Scheme Operation Example

# IV.   Performance Analysis

## A. Traditional PCF

In the traditional PCF scheme, all the packets have to be transmitted to LC through the dedicated channel $H_0$ one by one. The length of one Periodic Duration $T_{PCF}$ is:

$$T_{PCF} = SIFS + \frac{Poll}{C} + \delta + SIFS + \frac{X + Header}{C} + \delta,$$

where $C$ and $\delta$ denote the transmission rate and propagation delay respectively. The $Poll$ and $Header$ denote the length of Poll message and MAC Header respectively as their names suggest.

If each of the $M$ SMs needs to report $L$ packets to the LC, then the time used for the payload transmission is given as:

$$T_{PCF}^1 = LM\frac{X}{C}.$$

On the other hand, the total time spent on finishing the entire reporting transmission is given as:

$$T_{PCF}^2 = \frac{Beacon}{C} + SIFS + LM \times T_{PCF} + \frac{Poll}{C},$$

where $Beacon$ is the length of the Beacon message. Therefore, the throughput for the traditional PCF, $S_{PCF}$ follows:

$$S_{PCF} = T_{PCF}^1 \Big/ T_{PCF}^2.$$

## B. PCF with Cognitive Radio (CR-PCF)

As we mentioned above, the SIFS interval in the traditional PCF is replaced with $\tau_{Sense}$ and $\tau_{Switch}$ so as to guarantee the channel sensing and channel switch successfully. Therefore, the length of the Periodic Duration $T_{CRPCF}$ is modified as:

$$T_{CRPCF} = \tau_{Sense} + \frac{Poll}{C} + \delta + \tau_{Switch} + \frac{X + Header}{C} + \delta,$$

Since all the channels are mutually independent, thus whenever the LC senses the channels, there are average $n = N\gamma$ channels which are not occupied by PUs and able to be used for LC-SM communication. For each of the $n$ channels, due to the memoryless property of exponential distribution, the probability $p$ that LC-SM communication will not be interrupted by the PU during the Periodic Duration is given as:

$$p = Pr(T_{OFF} < y) = 1 - e^{-\lambda y},$$

where $y = T_{CRPCF} - \tau_{Sense}$ and $\lambda = \frac{1}{Z_{OFF}}$. Therefore, the expected number of available channels per Periodic Duration is $ne^{-\lambda y}$ and accordingly, the number of total Periodic Durations is approximately $\frac{LM}{ne^{-\lambda y}+1}$. Since in each Periodic Duration, there are $n$ SMs switching channels twice, thus the total number of channel switches $O$ is given as:

$$O = \frac{LM}{ne^{-\lambda y} + 1} 2n.$$

Then the total time for the entire transmission is given as:

$$T^2_{CRPCF} = \frac{Beacon}{C} + \tau_{Sense} + \frac{LM}{ne^{-\lambda y} + 1} T_{CRPCF} + \frac{Poll}{C},$$

As a result, the throughput for the CR-PCF, $S_{CRPCF}$ follows:

$$S_{PCF} = T^1_{CRPCF} \big/ T^2_{CRPCF} = LMX \big/ CT^2_{CRPCF}.$$

Furthermore, it is noted that extending the length of payload $X$ will both increase the good throughput and make its transmission more vulnerable to the interruption of PUs, which will decrease the throughput. Therefore, given the traffic behavior of PUs, it is valuable to calculate the optimal $\bar{X}$ so as to maximize the throughput $S$, which is derived as:

$$\bar{X} := \frac{dS_{CRPCF}}{dX} = 0$$

# V. Performance Analysis

In this section, we compare the performance of traditional PCF scheme and CR-PCF. All the results are presented based on the numerical equations derived above and the simulation run by NS-3. The default values of the parameters used in the section are summarized in the Fig.7. when they are not regarded as the variables in X-axis.

| Parameter | Value | Parameter | Value |
|---|---|---|---|
| Beacon | 100 byte | Poll-ACK Message | 50 byte |
| Payload $X$ | 1024 byte | Header | 50 byte |
| SIFS | 16 $\mu s$ | $\delta$ | 1 $\mu s$ |
| $\tau_{Sense}$ | 300 $\mu s$ | $\tau_{Switch}$ | 120 $\mu s$ |
| $Z_{ON}$ | 8000 $\mu s$ | $Z_{OFF}$ | 25000 $\mu s$ |
| Transmission Rate $C$ | 1 $Mbps$ | Number of SMs $M$ | 600 |
| Number of Ch. $N$ | 15 | Number of Packets per SM $L$ | 15 |

Fig.7　Parameter List for Numerical and Simulation Analysis

A. Performance with respect to Number of SM *M* and Number of Packets per SM *L*

As shown in the Fig.8, 9 the throughputs of both schemes are almost constant, which is because the throughput easily gets saturated as *L* and *M* are large enough. We may see that the throughput of the CR-PCF is much better than that of traditional PCF. This gain comes from the parallel transmissions over the available channels without the interference of PUs. On the other hand, the overall number of channel switches is linearly increased as we expected based on the equation in Section IV. It is also noted that, the simulation results are quite accurate compared to the numerical results especially as *LM* is large enough, which exactly matches the case of LC-SM communication.

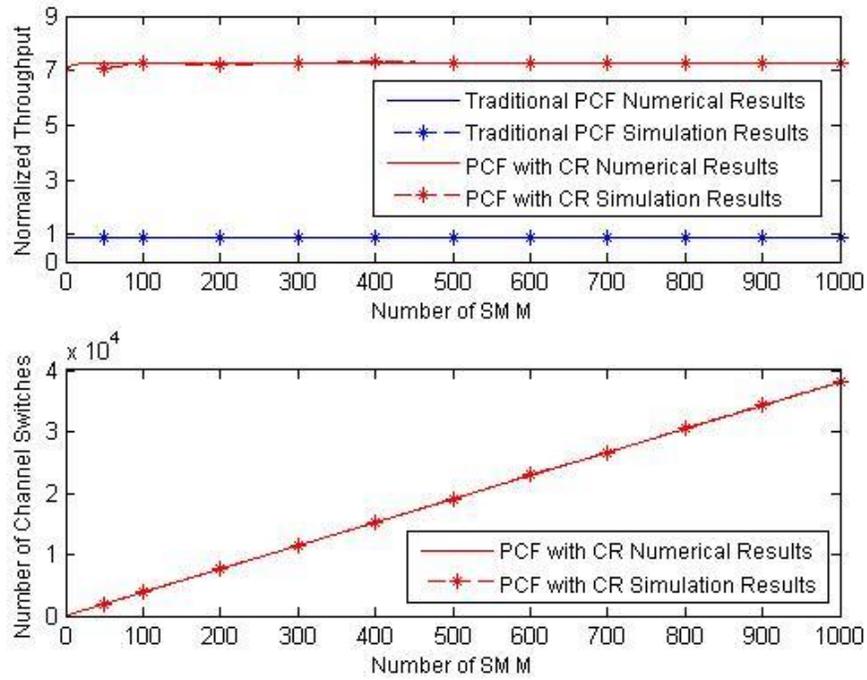

Fig.8    Performance Vs. Number of SM *M*

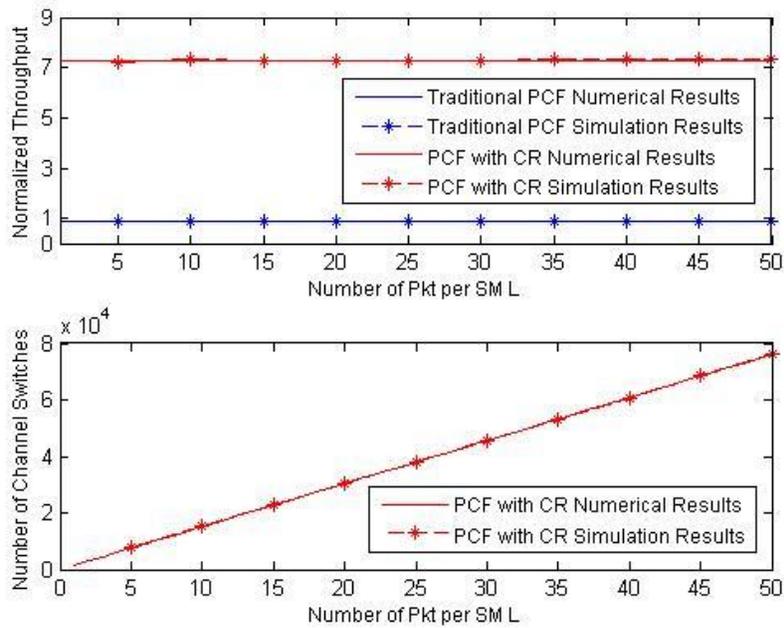

Fig.9    Performance Vs. Number of Packets per SM *L*

## B. Performance with respect to Number of Channels *N*

As shown in the Fig.10, the throughput of the CR-PCF is approximately linearly increased as we give more available channels to the network. It is quite reasonable that more available channels will lead to more parallel pipes to be used for transmissions. On the other hand, the number of channel switches rises dramatically at first because there are more SMs have chance to switch to non-dedicated channels. After that, it gets saturated when the number of channels is large enough. This is because the rounds of Periodic Duration is decreased as more channels are available.

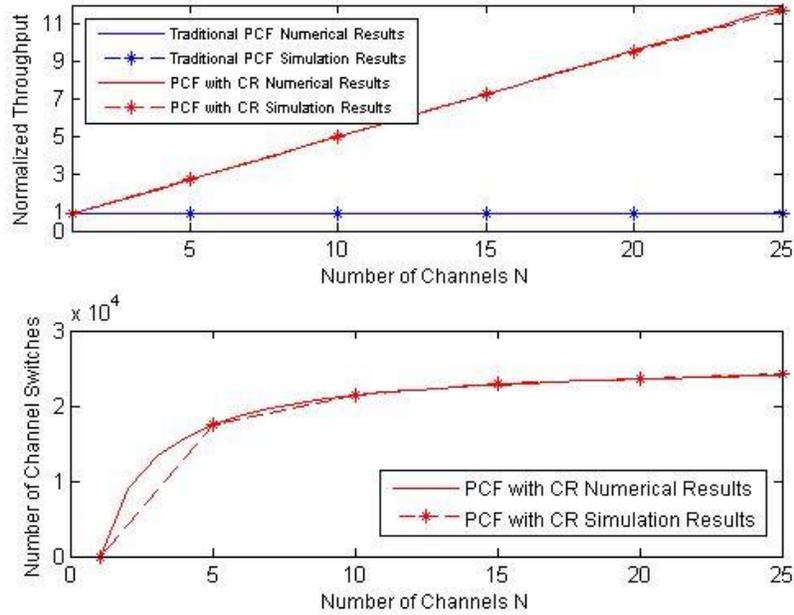

Fig.10  Performance Vs. Number of Channels $N$

## C. Performance with respect to Length of Payload $X$

As shown in the Fig.11, the throughput of traditional PCF scheme is getting saturated as length of packets $X$ becomes large enough, which is because the ratio between Payload and Header is getting larger and the time portion wasted on the Header transmission also shrinks. Similarly, the throughput of the CR-PCF rises at first stage. However, it starts to drop after the optimal length of Payload because the longer payload will make its transmission more vulnerable to the interruption of PU. Then it will lead to more retransmissions and decrease the throughput. As a result, there will be more rounds of Periodic Duration required, which also increases the overall number of channel switches as shown in the Fig.11.

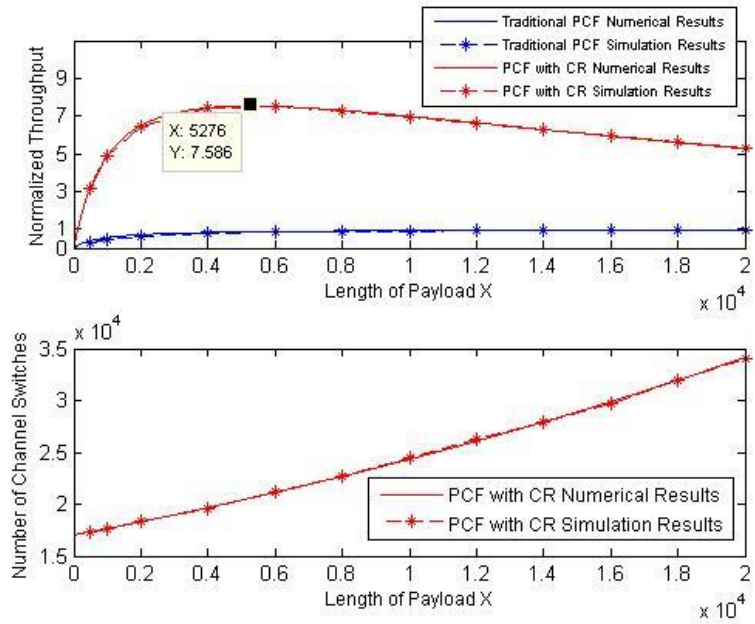

Fig.11  Performance Vs. Length of Payload *X*

### D. Performance with respect to the long-run proportion of channel off $\gamma$

As shown in the Fig.12, the throughput of the CR-PCF is increasing as $\gamma$ becomes large. This is because the less the PUs occupy the channels, the more chances that the SMs can use the available channels to transmit their data. As a consequence, both of the rounds of Periodic Duration required and resulting overall number of channel switches decrease.

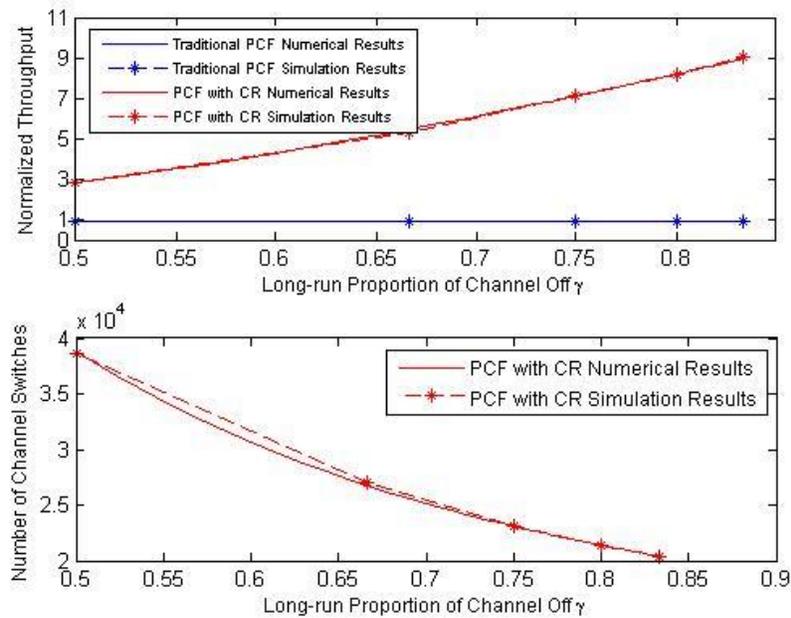

Fig.12  Performance Vs. Long-run Proportion of Channel Off $\gamma$

# VI. Conclusion

In this paper, in order to support the application of periodic energy consumption report in a densely populated network, Smart Metering Network, we propose a modified PCF MAC protocol with the combination of Cognitive Radio technology, in which the Smart Meters and Local Collector may communicate with each other through the white space when PUs are not present. We conduct the throughput analysis for this proposed scheme. Furthermore, the numerical results and simulation results through NS-3 show that the PCF scheme with CR significantly outperform the traditional one.

# Reference

[1] V.C.Gungor, D.Sahin, T.Kocak, S.Ergut, C.Buccella, C.Cecati, and G.P.Hancke, "Smart Grid Technologies: Communication Technologies and Standards," IEEE Transactions on Industrial Informatics, vol.7, no.4, Nov. 2011.
[2] T.Khalifa, K.Naik, and A.Nayak, "A Survey of Communication Protocols for Automatic Meter Reading Applications," IEEE Communications Survey and Tutorials, vol.13, no.2, Second Quarter, 2011.
[3] Y.Yang, and S. Roy, "PMU Deployment for Three-Phase Optimal State Estimation Performance," IEEE International Conference on Smart Grid Communications, 2013, Vancouver, Canada.
[4] Y.Yang, and S. Roy, "PMU Deployment for Optimal State Estimation Performance," IEEE GLOBECOM 2012, Anaheim, CA, USA.
[5] IEEE 802.11ah TG, "Specification Framework for TGah," IEEE 802.11-11/1137r15, May 2013.
[6] C.Lima, "Smart Grid Communications Logical Reference Architecture," IEEE P2030-09-0110-00-0011, Oct. 2009.
[7] Y.Yang, and S.Roy, "Grouping Based MAC Protocols for EV Charging Data Transmission in Smart Metering Network," IEEE Journal of Selected Area on Communications, vol.32, no 7, July 2014.
[8] National Electrical Manufacturers Association "American National Standard for Utility Industry End Device Data Tables," 2008.
[9] Y.Yang, X.Wang, and X.Cai, "On the Number of Relays for Orthogonalize-and-Forward Relaying," IEEE WCSP 2011, Nanjing, China.
[10] W.Wang, Y.Xu, and M.Khanna, "A Survey on the Communication Architectures in Smart Grid," Computer Networks, vol.55, no.16, Oct.2011.
[11] A.A.Sreesha, S.Somal, and I.Lu, "Cognitive Radio based Wireless Sensor Network Architecture for Smart Grid Utility," 2011 IEEE Systems, Applications and Technology Conference, May, 2011.
[12] A. Ghassemi, S. Bavarian, and L.Lampe, "Cognitive Radio for smart Grid Communications," 2010 IEEE International Conference on Smart Grid Communications, Oct. 2010.
[13] F.Liu, J.Wang, Y.Han, and P.Han, "Cognitive Radio Networks for Smart Grid Communications," 2013 Asian Control Conference, Jun. 2013.
[14] R.Yu, Y.Zhang, S.Gjessing, C.Yuen, S.Xie, and M.Guizani, "Cognitive Radio based Hierarchical Communications Infrastructure for Smart Grid," IEEE Magazine on Network, vol.25, no.5, 2011.
[15] Research Report of SBI Research, http://www.sbireports.com/Smart-Grid-Utility-2496610/.
[16] IEEE 802.11ah TG, "Potential Compromise for 802.11ah Use Case Document," IEEE 802.11-11/0457r0, Mar. 2011.
[17] Y. Yang, "Contributions to smart metering protocol design and data analytics," Ph.D. dissertation, University of Washington.
[18] Y.Lin, S.Liu, G.Qiao, and Y.Yang, "OFDM Demodulation Using Virtual Time Reversal Processing in Underwater Acoustic Communications", Journal of Computation Acoustics, vol.23, no.4, 2015.